\documentclass[aps,twocolumn,showpacs,superscriptaddress]{revtex4}
\usepackage{graphicx}

\begin{document}

\title{Wave function collapses in
a single spin magnetic resonance force microscopy}
\author{G.P. Berman}
\affiliation{Theoretical Division,MS B213, Los Alamos National
Laboratory, Los Alamos, NM 87545}

\author{F. Borgonovi}
\affiliation{Dipartimento di Matematica e Fisica, Universit\`a
Cattolica, via Musei 41, 25121 Brescia, Italy} \affiliation{INFM,
Unit\`a di Brescia and INFN, Sezione di Pavia, Italy}

\author{V.I. Tsifrinovich}
\affiliation{IDS Department, Polytechnic University, Six Metrotech
Center, Brooklyn, New York 11201}

\date{\today}

\begin{abstract}

{We study the effects of wave function collapses in the
oscillating cantilever driven adiabatic reversals (OSCAR) magnetic
resonance force microscopy (MRFM) technique. The quantum dynamics
of the cantilever tip (CT) and the spin is analyzed and simulated
taking into account the magnetic noise on the spin. The deviation
of the spin from the direction of the effective magnetic field
causes a measurable shift of the frequency of the CT oscillations.
We show that the experimental study of this shift can reveal the
information about the average time interval between the
consecutive collapses of the wave function.}

\end{abstract}

\maketitle

The oscillating cantilever driven adiabatic reversals (OSCAR)
technique appears to be the shortest way to a single spin
detection in magnetic resonance force microscopy (MRFM) \cite{1}.
In this technique the cantilever tip (CT) vibrations in
combination with a radio-frequency ({\it rf}) resonance field
causes adiabatic reversals of the effective magnetic field. Spins
of the sample follow the effective magnetic field causing a small
shift of the CT frequency, which is to be measured. The
quasiclassical theory of the OSCAR technique has been developed in
\cite{2}-\cite{6}. The quantum theory of a single-spin measurement
in OSCAR MRFM was presented in \cite{7,8}.

It was shown in \cite{7,8} that the CT frequency can take two
values corresponding to two possible directions of the spin
relative to the direction of the effective magnetic field, ${\vec
B}_{eff}$. If the spin points initially in (or opposite to) the
direction of ${\vec B}_{eff}$, then the CT has a definite
trajectory with the corresponding  positive (negative) frequency
shift. In the general case, the Schr\"odinger dynamics describes a
superposition of two possible trajectories - the Schr\"odinger cat
state. It was also shown in \cite{7,8} that the interaction
between the CT and the environment causes two effects. The first
one is a rapid decoherence: the Schor\"odinger cat state
transforms into a statistical mixture of two possible
trajectories. Physically this means that the CT quickly selects
one of two possible trajectories, even if initially the spin does
not have a definite direction relative to the direction of ${\vec
B}_{eff}$. The second effect is an ordinary thermal diffusion of
the CT trajectory.

What was not taken into account in \cite{7,8} was the effect of
the direct interaction between the spin and the environment. In
general, for an MRFM system one should consider two environments:
one for the CT and the other for the spin. If the initial spin
wave function describes a superposition of two spin directions
relative to  ${\vec B}_{eff}$, then the spin generates two
trajectories for the CT. The cantilever is a quasiclassical device
which measures the spin projection relative to the direction of
${\vec B}_{eff}$. The CT environment collapses the CT-spin wave
function selecting only one CT trajectory and a definite direction
of the spin relative to ${\vec B}_{eff}$. This situation is
similar to the Stern-Gerlach effect, but for a time-dependent
${\vec B}_{eff}$. The direct interaction of the spin with its
environment is extremely weak in comparison to the interaction
between the CT and its environment. However, this weak interaction
causes a deviation of the spin from the direction of ${\vec
B}_{eff}$ after a collapse of the wave function. In turn, this
deviation generates two CT trajectories. This scenario occurs
again and again in the OSCAR MRFM. Normally, a collapse ``forces''
the spin back to its initial (after the previous collapse)
direction relative to ${\vec B}_{eff}$.  But sometimes this
collapse pushes the spin to the opposite direction, revealing the
quantum jump - a sharp change of the spin direction and CT
trajectory. It was shown in \cite{4}-\cite{6} that the main source
of the magnetic noise on the spin was associated with the
cantilever modes whose frequencies were close to the Rabi
frequency.

There are two basic problems associated with the single spin OSCAR
MRFM. The first problem is the theoretical description of the
statistical properties of quantum jumps. Unfortunately, the direct
simulation of quantum jumps consumes too much computer time to be
implemented. Recently, we have considered a simplified model which
describes the statistical properties of quantum jumps \cite{9}.
The second problem is more sophisticated: What is the
characteristic time interval between two consecutive collapses of
the the wave function? This Letter discusses the second problem.

The CT position and momentum have finite quantum uncertainties.
Thus, when the spin direction deviates from the direction of
${\vec B}_{eff}$, the collapse does not occur instantly. During a
finite time interval, the spin and the CT are entangled. The spin
does not have a definite direction, and the CT does not have a
definite trajectory. The dynamics of the CT-spin system on the
time scale less than the time interval between two consecutive
collapses can be described by the Schr\"odinger equation. During
this time, the average CT frequency shift is expected to be
smaller (in absolute value) than the frequency shift corresponding
to the definite direction of the spin relative to ${\vec
B}_{eff}$. We show that the experimental study of this effect can
reveal the answer to a fundamental problem of quantum dynamics: At
what time does the collapse of the wave function occur if the
quasiclassical trajectories are not well separated?

Indeed, if the quasiclassical trajectories are initially well
separated (the Schr\"odinger cat state) then the characteristic
time of the collapse is the decoherence time, i.e. the time of
vanishing of the non-diagonal peaks of the density matrix (the
non-diagonal peaks describe the quantum correlation between two
trajectories when these trajectories coexist during the same time
interval \cite{10,11}). However, the Schr\"odinger cat state is a
specific bizarre phenomenon in the macroscopic world. Namely, in a
typical situation the collapse of the wave function occurs long
before a well defined separation develops between the two
quasiclassical trajectories. There are a few simple cases, for
which the exact solutions of the master equation have been
obtained (see, for example, \cite{11}). The exact solution
describes a generation of the two quasiclassical trajectories,
their decoherence, and a thermal diffusion. Before the two
quasiclassical trajectories are well separated, the exact solution
describes the complicated dynamics of the density matrix elements.
It is not clear if the master equation is capable to describe the
wave function collapse when the two trajectories are not well
separated. Even if the collapse time for this case is ``hidden''
in the solution of the master equation, we still do not know how
to extract it analytically and numerically. Only experiments could
resolve this fundamental problem. We show that OSCAR MRFM could
become one of these experiments.

\begin{figure}
\includegraphics[scale=0.36]{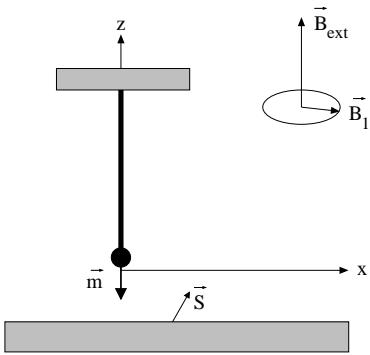}
\caption{ Cantilever set-up. $\vec{B}_{ext}$ is the external
static magnetic field, $\vec{B}_1$ the rotating magnetic field,
$\vec{m}$ the magnetic moment of the ferromagnetic particle glued
on the cantilever tip, $\vec{S}$ the spin to be measured. }
\label{des}
\end{figure}

The OSCAR MRFM setup considered here is shown in Fig. \ref{des}.

The dimensionless Hamiltonian for the OSCAR MRFM in the rotating
system of coordinates is taken as
\begin{equation}
{\cal H}=(p_x^2+x^2)/2+\varepsilon S_x+2\eta x S_z, \label{ham1}
\end{equation}
where we use the quantum units of the momentum $P_0=\hbar/X_0$ and
the coordinate $X_0=\sqrt{\hbar\omega_c/k_c}$. Here $\omega_c$ is
the cantilever frequency, $k_c$ is the cantilever effective spring
constant, $\varepsilon=\gamma B_1/\omega_c$, $\gamma$ is the
magnitude of the electron gyromagnetic ratio and
$\eta=(1/2)(\gamma X_0/\omega_c)G$, where $G = \partial
B_{z}/\partial x$, is the magnetic field gradient at the spin
location. The first term in (\ref{ham1}) describes the unperturbed
oscillations of the CT, the second term describes the interaction
between the spin and the resonant {\it rf} field, and the last
term describes the interaction between the spin and the CT. To
take into consideration the magnetic noise on the spin caused by
the thermal CT vibrations we add a random term $\Delta(t)S_z$ to
the Hamiltonian (\ref{ham1}).

The wave function of the system is assumed to have the form
\begin{equation}
\Psi=u_\alpha(x,\tau){\bf{\alpha}}+u_\beta(x,\tau){\bf{\beta}},
\label{psi4}
\end{equation}
where ${\bf{\alpha}}$ and ${\bf{\beta}}$ are the basis spin
functions in the $S_z$ representation corresponding to the values
$S_z=\pm 1/2$, and $\tau=\omega_ct$  is the dimensionless time.
The Schr\"odinger equation splits into two coupled equations for
$u_\alpha(x,\tau)$ and $u_\beta(x,\tau)$. If these two functions
are identical (up to a constant factor) then the wave function can
be represented by a product of the CT and spin wave functions. In
this case the average spin $\langle {\vec S}\rangle$  has a
magnitude 1/2. In the general case, the spin is entangled with the
CT, and the average spin is smaller than 1/2. In our estimates we
will use the following parameters from experiment \cite{1}:
\begin{equation}
\begin{array}{lll}
&\omega_c/2\pi=6.6 kHz,~k_c=600 \mu N/m,~B_1=300\mu T,\\
&G=4.3\times 10^5 T/m,~X_m=10 nm,~T=200 mK,\\
\label{P}
\end{array}
\end{equation}
where $X_m$ is the amplitude of the CT. Using these values, we
obtain the following values of parameters for our model
\begin{equation}
\begin{array}{lll}
X_0 &= 85 fm, ~P_0 = 1.2\times 10^{-21} Ns, ~\eta = 0.078,\\
x_m &= X_m/X_0 = 1.2\times 10^5.
\end{array}
\label{A}
\end{equation}
The relative frequency shift
$\delta\omega_0=|\Delta\omega_c/\omega_c|$ of the cantilever
vibrations can be estimated to be \cite{6,7}:
\begin{equation}
\delta\omega_0=2S\eta^2/(2\eta^2x^2_m+\varepsilon^2)^{1/2}=4.7\times
10^{-7}. \label{B}
\end{equation}
(We insert the factor $2S$ for future discussion.)

Suppose that initially the spin points opposite to the direction
of ${\vec B}_{eff}$. The frequency shift of the CT vibrations is
then $-\delta\omega_0$. The magnetic noise acting on the spin
causes a deviation of the spin direction from the direction of the
effective magnetic field. Thus, it produces two trajectories of
the CT with the frequencies $\pm\delta\omega_0$. Because of the
quantum uncertainty of the CT position during the finite time
(which we call the collapse time $\tau_{coll}$), the wave function
of the CT describes a single peak with an absolute value of the
frequency shift less than $\delta\omega_0$. The collapse of the
wave function changes the frequency shift to the value
$-\delta\omega_0$ (or, sometimes, $\delta\omega_0$, in the case of
a quantum jump).

We simulate the quantum dynamics between two consecutive collapses
of the wave function of the system. In our simplified model the
magnetic noise $\Delta(t)$ takes consecutively two values,
$\pm\Delta_0$. The time interval between two consecutive ``kicks''
of the function $\Delta(t)$ was taken randomly from the interval
$(-3\tau_R/4,5\tau_R/4)$, where $\tau_R=2\pi/\varepsilon$ is the
dimensionless Rabi period. (In a more advanced theory the
characteristics of the magnetic noise should be derived from the
parameters of the thermal CT vibrations.) The initial wave
function is assumed to be a direct product of the CT and spin wave
functions. The initial state of the CT is a coherent state
\begin{equation}
|\alpha_0\rangle=\pi^{1/4}\sum_{n}2^{n/2}\alpha_0^nH_n(x)\exp\{-(x^2+|\alpha_0|^2)/2\},
\label{coh5}
\end{equation}
where $H_n(x)$ is a Hermite polynomial, and
$\alpha_0={{1}\over{\sqrt{2}}}(x_0+p_0)$; here $x_0$ and $p_0$ are
the quantum mechanical averages of $x$ and $p$ at $\tau=0$. The
initial direction of the spin is taken to be opposite to the
direction of the effective magnetic field $\vec B_{eff}=\vec
i\varepsilon +2\vec k\eta x_0$, where $\vec i$ and $\vec k$ are
the unit vectors in the positive $x$- and $z$-directions.

In our numerical simulations we expand the functions
$u_\alpha(x,\tau)$ and $u_\beta(x,\tau)$ in (\ref{psi4}) over 400
eigenfunctions of the unperturbed oscillator Hamiltonian. During
the time interval between two consecutive ``kicks'' of the noise
function $\Delta(t)$ we have a time-independent Hamiltonian. Thus,
we find the evolution of the wave function by diagonalizing the
$800\times 800$ matrix and taking into consideration the initial
conditions after each ``kick''. The output of our simulations is
the time interval $\Delta\tau_j=\tau_j-\tau_{j-1}$ between two
consecutive returns to the origin for the average value $\langle
x\rangle$: $\langle x(\tau_{j-1})\rangle=0$ and $\langle
x(\tau_j)\rangle=0$. To save computational time, we have used the
values of parameters $\varepsilon=10$, $\eta=0.3$, $p_0=0$,
$x_0=13$. These values provide a relative large frequency shift of
the CT oscillations \cite{7,8}
\begin{equation}
\delta\omega_0\approx 7.9\times 10^{-3}. \label{eta8}
\end{equation}

\begin{figure}
\includegraphics[scale=0.3]{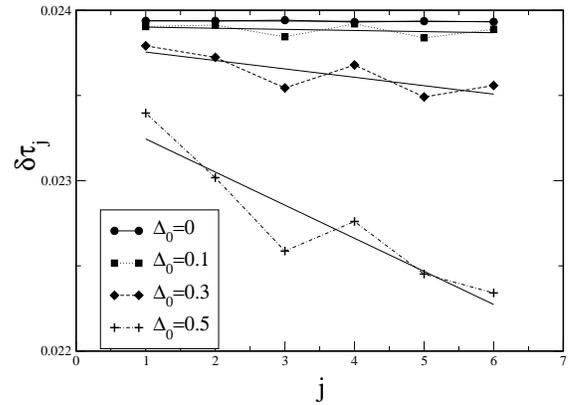}
\caption{ Deviation of $\Delta\tau_j$ from the unperturbed half
period of the CT oscillations $\pi$, as a function of the number
of half periods $j$, for different $\Delta_0$ as indicated in the
legend. Solid lines are the standard linear fits. On the vertical
axis $\delta\tau_j=|\Delta\tau_j-\pi|$.} \label{r1}
\end{figure}

The results of our simulations are shown in Fig. \ref{r1}, which
demonstrates the deviation of $\Delta\tau_j$ from the unperturbed
half-period of the CT oscillations $\pi$. We introduce
$\delta\tau_j=|\Delta\tau_j-\pi|$. With no magnetic noise
$(\Delta_0=0)$ the deviation $\delta\tau_j$ does not change with
time
\begin{equation}
\delta\tau_j=\delta\tau_0=\pi/\delta\omega_0\approx 0.025.
\label{del9}
\end{equation}

If $\Delta_0\not=0$, the value of $\delta\tau_j$ decreases with
time until the collapse of the wave function destroys the
Schr\"odinger cat state. (Fig. 2 demonstrates the case for which
the time interval between the collapses equals six half-periods of
the CT vibrations).

We will define the ``effective frequency'' $\omega_j$ for each
half-period of the CT vibrations: $\omega_j=\pi/\Delta\tau_j$.
Next, we introduce the effective frequency shift
$\delta\omega_j=\pi/ \delta \tau_j$.  From Fig. 2 we can find the
average effective frequency shift $\langle\delta\omega\rangle$
between two collapses of the wave function.

If our model of the noise were the adequate one, then from the
experimentally measured quantity $\langle\delta\omega\rangle$ we
could determine the time interval between two collapses
$\tau_{coll}$. (It is clear from Fig. 2 that
$\langle\delta\omega\rangle$ is directly related to the value of
$\tau_{coll}$ for the assumed magnetic noise parameters.)
Certainly, in a real situation this opportunity does exist if the
average frequency shift $\langle\delta\omega\rangle$ is
significantly smaller than the expected shift $\delta\omega_0$. We
will call such situation ``the case of the strong noise''.

Note, that a decrease of the frequency shift may be interpreted as
an effective decrease of the spin $\delta S$. Using Eq.(\ref{B}) we obtain
\begin{equation}
\delta
S=(\langle\delta\omega\rangle-\delta\omega_0)
(2\eta^2x_m^2+\varepsilon^2)^{1/2}/(2\eta^2).
\label{D}
\end{equation}

Previously we have shown that in the case of strong noise, an
experimentalist could determine the time interval $\tau_{coll}$
between two consecutive collapses of the wave function by
measuring the decrease of the frequency shift of the CT
vibrations.  Here we propose a special experiment which allows one
to determine the time $\tau_{coll}$  for the case of weak noise in
which the average frequency shift between two collapses is close
to $\delta\omega_0$.

The problem is the following. Even a very weak noise generates a
second CT trajectory with a frequency shift opposite to that of
the first trajectory. Thus, two trajectories tend to separate at
the same rate for any magnetic noise. Correspondingly, the time
interval between two collapses is expected to be approximately the
same for any level of magnetic noise. However, in the case of weak
noise, the probability of the second trajectory is small, so its
contribution to the average frequency shift becomes negligible.

To overcome this obstacle, we propose an artificial change of the
frequency shift using the ``interrupted OSCAR technique'' recently
implemented in \cite{1}. In \cite{1}, the {\it rf} field is turned
off for a time interval equal to half of the CT period, which is
equivalent to the application of the $\pi$-pulse which changes the
direction of the spin relative to the effective magnetic field. We
propose to turn off the {\it rf} field for the duration of the
quarter of the CT period, which is equivalent to the application
of the $\pi/2$-pulse. Suppose that initially the spin is parallel
to the effective magnetic field, ${\vec B}_{eff}$. If we apply a
``$\pi/2$-pulse'', the spin will become perpendicular to ${\vec
B}_{eff}$. Thus, we have two CT trajectories, each with the same
probability. Before these two trajectories are separated, the CT
will oscillate with the unperturbed frequency, which is equal to
one in our dimensionless units. After the collapse the frequency
shift is $\pm \delta\omega_0$, with equal probabilities. Thus,
using a ``$\pi/2$-pulse'' we can achieve a maximum possible
reduction of the frequency shift. If we apply a periodic sequence
of ``$\pi/2$-pulses''  with the period $\tau_p$
($\tau_p>\tau_{coll}$), then the average frequency shift
$\langle\delta\omega\rangle$ is
\begin{equation}
\langle\delta\omega\rangle=\delta\omega_0(\tau_p-\tau_{coll})/\tau_p.
\label{E}
\end{equation}
Manipulating $\tau_p$ one can achieve a significant decrease of
$\langle\delta\omega\rangle$ in comparison with $\delta\omega_0$.
Using Eq. (\ref{E}) one can determine the collapse time from the
experimental value $\langle\delta\omega\rangle$. Thus, the
collapse time can be measured for the case of weak magnetic noise.
Finally, based on quasiclassical theory \cite{5} we will estimate
the reduction of the average frequency shift caused by the noise
for the experimental conditions \cite{1}. The amplitude of the
thermal CT vibrations near the Rabi frequency can be estimated to
be:
\begin{equation}
a_T=(\omega_c/\omega_R)(k_BT/2k_c)^{1/2}=38 fm. \label{F}
\end{equation}
The square of the characteristic spin deviation during a single
reversal (half of the CT period) is
\begin{equation}
(\Delta\theta_1)^2\approx 3.4Ga^2_T/(\omega_cX_m)=9\times 10^{-7}.
\label{G}
\end{equation}
We have no idea about the order of the collapse time
$\tau_{coll}$. If we assume that the collapse occurs when the
separation between the two trajectories with the frequencies
$1\pm\delta\omega_0$ is equal to $1/2$ (the quantum uncertainty of
the CT position in the coherent state is $\langle(\Delta
x)^2\rangle=1/2$), then we obtain for $\tau_{coll}$
\begin{equation}
\tau_{coll}\sin\tau_{coll}\approx 1/(4x_m\delta\omega_0)\approx
4.4. \label{H}
\end{equation}
It follows from Eq.(\ref{H}) that the wave function collapses
during the second period of the CT vibrations. If we estimate the
probabilities of the two CT trajectories as $P_1\sim 1-
(\Delta\theta_1)^2$ (for the trajectory with the initial frequency
shift) and $P_2\sim(\Delta\theta_1)^2$ (for the trajectory with
the opposite frequency shift), then the average CT frequency shift
can be estimated to be
\begin{equation}
\langle\delta\omega\rangle=\delta\omega_0(P_1-P_2)=\delta\omega_0(1-2(\Delta\theta_1)^2).
\label{K}
\end{equation}
This estimated reduction of the CT frequency shift is clearly
negligible.

Consider the opposite extreme case. Suppose that the collapse
occurs when the separation between the two trajectories is of the
order of the thermal CT fluctuations $(k_BT/k_c)^{1/2}\approx
150$pm or $1760$ in dimensionless units (as before, we used the
values of parameters in (3)). In this case, the time interval
between the two consecutive collapses $\tau_{coll}$ is of the
order of $10^4$ periods of the CT oscillations. During this time,
the characteristic spin deviation is $(\Delta\theta)^2\sim 0.02$.
Then, we have for the probability $P_1\approx
(\Delta\theta)^2/4\approx 5\times 10^{-3}$. The average frequency
shift is $\langle\delta\omega\rangle\approx
\delta\omega_0(1-10^{-2})$. One can see that even in this extreme
case the reduction of the frequency shift is expected to be small.

Thus, the experimental conditions in \cite{1} probably correspond
to the case of the weak noise. In such a situation the collapse
time could be measured using the periodic sequence of
``$\pi/2$-pulses'' described earlier.

 We have demonstrated a
procedure for measuring the mysterious collapse time in the OSCAR
MRFM technique. We simulated the quantum dynamics of the spin-CT
system. Unlike the previous studies of the quantum dynamics we
took into consideration the direct interaction between the spin
and the environment (the magnetic noise). This noise causes (i) a
deviation of the spin from the direction of the effective magnetic
field and (ii) entanglement between the spin and the CT. For the
case of weak magnetic noise, the same effect can be achieved the
effective ``$\pi/2$-pulses'' with the ``interrupted OSCAR''
technique. The spin-CT entanglement influences the frequency of
the CT oscillations before the wave function collapse takes place.
This effect can be described as an effective decrease of the
single spin magnitude. We demonstrated that the experimental
measurement of the OSCAR MRFM frequency shift could reveal
information about the time interval between two consecutive
collapses of the wave function.

We thank D. Rugar and G.D. Doolen for discussions. This work was
supported by the Department of Energy (DOE) under Contract No.
W-7405-ENG-36, by the Defense Advanced Research Projects Agency
(DARPA), by the National Security Agency (NSA), and by the
Advanced Research and Development Activity (ARDA).

\end{document}